\documentclass[%
superscriptaddress,
preprintnumbers,
nofootinbib,
 amsmath,amssymb,
 aps,
]{revtex4-2}

\usepackage{graphicx}
\usepackage{subcaption}
\usepackage[justification=raggedright]{caption}
\usepackage{dcolumn}
\usepackage{bm}
\usepackage{hyperref}
\hypersetup{
  colorlinks   = true, 
  urlcolor     = blue, 
  linkcolor    = blue, 
  citecolor   = red 
}

\usepackage{parskip}
\usepackage{float}

\usepackage{slashed}

\def\be{\begin{equation}}
\def\ee{\end{equation}}
\def\bea{\begin{eqnarray}}
\def\eea{\end{eqnarray}}
\def\bear{\begin{array}}
\def\ear{\end{array}}
\def\bfig{\begin{figure}}
\def\efig{\end{figure}}
\def\bcen{\begin{center}}
\def\ecen{\end{center}}
\def\bi{\begin{itemize}}
\def\ei{\end{itemize}}

\def\D{\displaystyle}

\def\d16{d_{16}}

\usepackage{bm}

\usepackage{xcolor}
\usepackage{multirow}
\usepackage{slashed}
\usepackage{physics}
\usepackage{makecell}

\newcommand{\no}{\nonumber}

\newcommand{\mL}{\mathcal{L}}

\newcommand{\mO}{\mathcal{O}}

\newcommand{\mpi}{M_\pi}
\newcommand{\md}{\mathring{m}_\Delta}
\newcommand{\m}{\mathring{m}}
\newcommand{\g}{\mathring{g}_A}
\newcommand{\bfour}{\widetilde{b}_4}
\newcommand{\pthree}{\mO(p^3)}
\newcommand{\pfour}{\mO(p^4)}
\newcommand{\rA}{\langle r_A^2 \rangle}
\newcommand{\chidof}{\chi^2/{\rm dof}}

\newcommand{\e}{\epsilon} 

\begin{document}

\title{Extraction of the nucleon axial form factor from Lattice QCD\\ using NNLO chiral perturbation theory}

\author{Fernando Alvarado}
\email{f.alvarado@gsi.de}
\affiliation{GSI Helmholtzzentrum f\"ur Schwerionenforschung GmbH,\\
Planckstraße 1, 64291 Darmstadt, Germany}
\author{Luis Alvarez-Ruso}%
\email{Luis.Alvarez@ific.uv.es}
\affiliation{Instituto de Física Corpuscular (IFIC), CSIC‐Universitat de València, E-46980 Paterna, Valencia, Spain}

\date{\today}

\begin{abstract}
We calculate the nucleon axial form factor in relativistic chiral perturbation theory with $\Delta(1232)$ up to next-to-next-to-leading order (NNLO). Relevant low-energy constants are determined by fitting to recent lattice-QCD results at several pion masses, while accounting for the uncertainty associated with the truncation of the chiral expansion. We obtain a good description of the lattice data for momentum transfers up to $\sqrt{Q^2}\simeq0.6$ GeV and pion masses up to $\mpi\simeq400$ MeV. We find that the explicit inclusion of the $\Delta$ resonance is required to reproduce the lattice-QCD pion-mass dependence of the axial charge and axial radius, as well as the momentum dependence of the form factor. At the physical point we obtain $g_A=1.257\pm0.011$ and $\langle r_A^2\rangle=0.312\pm0.037~\mathrm{fm}^2$. Our analysis provides a model-independent and systematically improvable parametrization of the pion-mass and momentum dependence of the axial form factor, offering a framework for extrapolating lattice-QCD results to the physical point and for improving predictions of low-energy weak interactions involving nucleons.
\end{abstract}

\maketitle

\section{\label{sec:FAintro}Introduction}
The axial form factor, $F_A(Q^2)$, is a fundamental property of the nucleon. 
Given that the weak interaction is left-handed, \textit{i. e.} $V-A$, its effect on the nucleon depends not only on vector form factors (FF), which up to isospin-breaking corrections can be related to electromagnetic FF, but also on $F_A$ and on the pion-pole-dominated induced pseudoscalar form factor $F_P$. Therefore, the axial FF of the nucleon is a key ingredient in the description of weak processes. In particular, it shapes the neutrino-nucleon elastic and quasielastic scattering cross section, upon which neutrino oscillation experiments critically rely \cite{NuSTEC:2017hzk}.
 
The extraction of the nucleon axial form factor from experimental data is challenging and may be affected by experimental biases and model dependence. Pion electroproduction is one of the sources of information about this FF \cite{Nambu:1962lbq,Bernard:2001rs}, although the connection of this process and $F_A$ is model independent only at low $Q^2$ where chiral perturbation theory (ChPT) is applicable \cite{Bernard:1992ys}. $F_A$ can also be extracted from neutrino-deuteron scattering bubble chamber experiments performed through the 70es to the 90es of the XXth century~\cite{Bodek:2007ym,Meyer:2016oeg}, which suffered from considerable flux uncertainties and efficiency corrections, particularly at low $Q^2$. The recent measurement of the $\bar\nu_\mu \, p \rightarrow \mu^+ \, n$ cross section using the plastic-scintillator target of the MINERvA experiment, by subtracting the background from interactions on carbon \cite{MINERvA:2023avz}, has also allowed the extraction of $F_A$. Finally, at very low and fixed $Q^2$ it has also been obtained from weak muon capture in muonic hydrogen ($\mu H$) with ChPT input~\cite{Hill:2017wgb}.
  Any extraction of the FF requires that a certain functional form is fitted to the data. 
 Historically, the dipole parameterization has been extensively used, employing only one free parameter, the axial mass $M_A$, even though it is not physically justified and leads to underestimated uncertainties. 
  The so-called $z$-expansion, which relies on  the analytic structure of the FF dictated by QCD, 
  has also been widely employed, and is considered to yield more reliable results. 
   The $z$-expansion has a different number of free parameters depending on the power, $k_\mathrm{max}$, of $z$ at which the series is truncated. The choice of $k_\mathrm{max}$, priors and fit strategy has an impact on the extracted FF and its uncertainty. New z-expansion fits to determine $F_A(Q^2)$ from hydrogen and deuterium data have been reported in Ref. \cite{MINERvA:2025ygc}. A neural-network-based parametrization of the nucleon axial FF has also been proposed~\cite{Alvarez-Ruso:2018rdx}. Ultimately, with sufficiently precise experimental data, all parametrizations and fitting strategies are expected to yield consistent results.
   
Lattice QCD (LQCD) simulations provide another valuable source of information on nucleon properties, including FF, which can be extracted directly from the interactions dictated by the QCD Lagrangian. In the last ten years, the progress in this field has been significant~\cite{FlavourLatticeAveragingGroupFLAG:2024oxs} thanks to better computational resources, improved algorithms, and techniques to lower systematic errors. 
Nevertheless, LQCD computations are generally affected by several artifacts that lead to systematic errors: the lattice finite volume, $L$; the discretization spacing, $a$; the values of the light quark masses, typically higher than the physical ones. Reaching the physical limit from simulations affected by the aforementioned artifacts requires a chiral extrapolation to the continuum, which relies on a nontrivial parametrization of the relevant dependencies.

Therefore, theoretical input is essential to describe not only the $Q^2$ dependence of the axial FF, but also the effects of lattice artifacts. In this context, Chiral Perturbation Theory (ChPT), the low-energy effective field theory of QCD \cite{Gasser:1983yg, Gasser:1987rb}, offers a symmetry-based alternative to ad hoc parametrizations. Once ChPT low energy constants are fixed, it parametrizes the $F_A$ dependence on the momentum transfer and the light quark mass, which are expansion parameters of the theory. Chiral Perturbation Theory can also account for lattice-volume and lattice-spacing corrections in a systematic way~\cite{Beane:2003xv,Beane:2004rf,Hall:2025ytt,Hermsen:2025vds,Hermsen:2026tqc}. It is also useful for addressing contamination from excited states, a major issue in the baryon axial sector that has received considerable attention~\cite{Tiburzi:2009zp,Bar:2017kxh,RQCD:2019jai}.  

There is a clear synergy between ChPT and LQCD, as the latter constrains the low-energy constants (LECs), some of which are difficult to determine from experiment. These constants enter ChPT together with the quark-mass dependence, and lattice calculations at nonphysical quark masses help disentangle their individual valus. This provides valuable input for predicting other observables, such as the neutrino–nucleon elastic cross section at low momentum transfer \cite{Chen:2024kbh}.

These questions have motivated us to perform a global analysis of a combined set of recent LQCD data using NNLO ChPT as parametrization. The main goal is to describe the axial FF at low energies without relying on ad-hoc parametrizations. On top of that, we determine important values of the low-energy constants and keep track of the truncation error of ChPT, aiming for consistent results while testing the convergence of the chiral approach. 

As discussed below, previous works have employed the non-relativistic approximation or have computed contributions only up to the leading loop correction, $\pthree$ in the chiral expansion. Our aim here is to outperform the previous analyzes thanks to the relativistic EOMS renormalization~\cite{Fuchs:2003qc} and the explicit inclusion of $\Delta$ up to $\pfour$ (NNLO). In fact, the EOMS scheme guaranties that not only power counting but also analytic properties of loop functions are properly preserved. Moreover, the inclusion of $\pfour$ leads to a better estimation of the truncation uncertainty, given by differences between orders.

After introducing and motivating our study, the axial FF and the terms of the effective Lagrangian that contribute to it are presented in Sec. \ref{sec:FAinBChPT}, as well as the pertinent Feynman diagrams. The fit to LQCD results is described in Sec.~\ref{sec:FAfit}. We finish with a comparison with previous determinations of the axial radius of the nucleon, Sec. \ref{sec:rA}, followed by the concluding Sec. \ref{sec:FAconcl}.

\section{\label{sec:FAinBChPT}The nucleon axial form factor in relativistic BChPT}

In the isospin-symmetric case of two quark flavors, the matrix element of the axial isovector current       
\bea
A_\mu^a(x)=\bar{q}(x) \gamma_\mu\gamma_5\frac{\tau^a}{2} q(x)  \,,
\eea
with $q=(u,d)^T$ the quark-field doublet, taken between on-shell nucleon states of four-momenta $p$, $p'$ ($p^2 = p'^2 = m_N^2$), can be written as
\be\label{eq:axialmatrixelem}
\langle N(p')|A_\mu^a(0)| N(p)\rangle =  \bar{u}(p')\left[ F_A(t) \gamma_\mu+\frac{q_\mu}{2 m_N}F_P(t) \right]\gamma_5 \frac{\tau^a}{\D 2} u(p) \ , \ee
where $t\equiv -Q^2\equiv q^2$ and $q=p'-p$. The isovector nature of the current is manifest through the presence of the Pauli isospin matrices  $\tau^a$. $F_A$ and $F_P$ are the axial and induced pseudoscalar form factors, respectively. We study the axial FF, which can be expanded for small $t$ as
\begin{equation}\label{eq:FAq2expansion}
    F_A(t)=g_A\left[1+\frac{1}{6}\rA t+\mO(t^2) \right]\ ;
\end{equation}
$g_A\equiv F_A(0)$ is the axial charge and the axial radius squared is given by the slope at $t=0$: 
\begin{equation}
\label{eq:rAdeff}
\rA=\frac{6}{g_A} \dv{}{t}\left.F_A\right|_{t=0} \ .
\end{equation}

\subsection{Relevant terms of the effective Lagrangian}
\par
In this section, the terms in the effective Lagrangian, $\mL_{\rm{eff}}$, 
required for our calculation are introduced. We treat the $\Delta$ as an explicit degree of freedom with the so-called small scale expansion power counting of Refs.~\cite{Hemmert:1996xg,Hemmert:1997ye}. To compute $F_A$ up to NNLO we need
\bea
\mL_{\rm eff} \supset \mL_{\pi N}^{(1)}+\mL_{\pi \Delta}^{(1)}+\mL_{\pi N \Delta}^{(1)}+\mL_{\pi N}^{(2)}+\mL_{\pi N \Delta}^{(2)}+\mL_{ \pi \Delta}^{(2)}+\mL_{\pi N}^{(3)}\ ,
\eea
where superscripts indicate the chiral order and subscripts show the degrees of freedom present. 
The terms in $\mL_{\pi N}^{(1,3)}$, $\mL_{\pi \Delta}^{(1)}$ and $\mL_{\pi N \Delta}^{(1)}$ that are relevant for our study have been shown in Ref. \cite{Yao:2017fym}, using the same notation adopted here. In addition,  $\mO(p^2)$ contributions must be included to determine $g_A$ at NNLO. 
Following the notation of Ref.~\cite{Fettes:2000gb}, the required terms of the $\pi N$ Lagrangian are
\begin{equation}
    \mL_{\pi N}^{(2)} \supset  \bar{\Psi}\left(c_1\langle \chi_+\rangle-\frac{c_2}{8 \m^2}\left(\langle u_\mu u_\nu\rangle \{D^\mu,D^\nu\}+\rm{h. c. }\right)+\frac{c_3}{2}\langle u_\mu u^\mu\rangle+\frac{i c_4}{4}[u_\mu,u_\nu]\sigma^{\mu\nu}\right)\Psi,
\end{equation}
where $\Psi$ is the isospin doublet of the nucleon fields; $\{,\}$ ($[,]$) represent anticommutators (commutators), while $\langle \cdots \rangle $ denotes a trace over isospin. Keeping only isovector axial external fields $a_\mu \equiv a_\mu^a\tau^a/2$,
 \bea
 D_\mu&=&\partial_\mu+\Gamma_\mu\ ,\nonumber\\
  \Gamma_\mu&=&\frac{1}{2}[u^\dagger(\partial_\mu-i a_\mu)u+u(\partial_\mu+i a_\mu)u^\dagger]\ ,\nonumber\\
   u_\mu&=&i[u^\dagger(\partial_\mu-i a_\mu)u-u(\partial_\mu+i a_\mu)u^\dagger]\ ;
 \eea 
$\chi^+=u^\dagger\chi u^\dagger+u\chi^\dagger u$, with $\chi={\rm diag}(M_\pi^2,M_\pi^2)$. 

In $\mL_{\pi N\Delta}^{(2)}$, after redundant terms are eliminated from  Eq.~(67) of Ref.~\cite{Jiang:2017yda} (see also Sec.~3.1 of Ref.~\cite{Holmberg:2018dtv} and the Appendix of Ref.~\cite{Unal:2021byi}) only the following monomials 
\be
\mL_{\pi N\Delta}^{(2)} \supset \bar{\Psi}_{\alpha}^k\xi^{\frac{3}{2}}_{ki}\left\{-i\frac{b_1}{2} F_{\alpha \beta}^{+,i}\gamma^5 \gamma^\beta+i b_2  F_{\alpha \beta}^{-,i}\gamma^\beta + \frac{b_4}{2}\omega_\alpha^i\omega_\beta^j \gamma^\beta\gamma_5 \tau^j+\frac{b_5}{2}\omega_\alpha^j\omega_\beta^i \gamma^\beta\gamma_5 \tau^j \right\}\Psi+ \rm{h.c.}
\ee
contribute to the nucleon axial FF; $\Psi_\mu$ denotes the Rarita-Schwinger field of the $\Delta$ resonance. States $\{ \xi^{\frac{3}{2}}_{ij} \Psi_\mu^j \}_{i=1-3}$, given in terms of isospin-$3/2$ projectors $\xi^{\frac{3}{2}}_{ij} = \delta_{ij} - \tau_i \tau_j /3 $, are isospin doublets, whose explicit expressions in terms of the physical $\Delta$ states are derived, for example, in Appendix~A of Ref.~\cite{Hacker:2005fh}; $\omega^{\mu,i}=\frac{1}{2}\,\langle\tau^i u^\mu \rangle$ and $F^{\pm,i}_{\mu\nu}=\frac{1}{2}\langle F^{\pm}_{\mu\nu}\tau^i \rangle$ are isospin traces. In the latter, 
\be
F_{\mu \nu}^{\pm}= u^\dagger F_{\mu \nu}^R u\pm u F_{\mu \nu}^{L}u^\dagger \,
\ee
where
\bea
F^L_{\mu \nu} &=& \partial_\mu l_\nu-\partial_\nu l_\mu -i [l_\mu,l_\nu]\ ,\nonumber\\
F^R_{\mu \nu} &=& \partial_\mu r_\nu-\partial_\nu r_\mu -i [r_\mu,r_\nu] \,
\eea
with $r_\mu= -l_\mu=a_\mu $. 
Finally, also at $\mO(p^2)$~\cite{Jiang:2017yda,Siemens:2020vop}, 
\begin{equation}
\mL_{\pi\Delta}^{(2)}  \supset \bar{\Psi}^{i \mu}\xi^{\frac{3}{2}}_{ij}\left\{a_1 \langle\chi_+\rangle \delta^{jk}g_{\mu\nu}\right\}\xi^{\frac{3}{2}}_{kl}\Psi^{l\nu}.
\end{equation} 
introduces an $\mpi$ dependent correction to the $\Delta$ mass, in the same way as the term proportional to $c_1$ does for the nucleon mass. 

\subsection{\label{subsec:FATheo}Perturbative calculation}
\begin{figure}[h]
\begin{subfigure}[t]{0.12\textwidth}
    \centering\includegraphics[width=\textwidth]{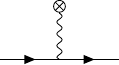}
    \caption{}
  \end{subfigure}\hfill
  \begin{subfigure}[t]{0.12\textwidth}
    \centering\includegraphics[width=\textwidth]{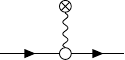}
   \caption{}
  \end{subfigure}\hfill
  \begin{subfigure}[t]{0.12\textwidth}
    \centering\includegraphics[width=\textwidth]{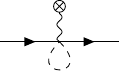}
   \caption{}
  \end{subfigure}\hfill
  \begin{subfigure}[t]{0.12\textwidth}
    \centering\includegraphics[width=\textwidth]{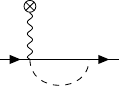}
    \caption{}
  \end{subfigure}\hfill
  \begin{subfigure}[t]{0.12\textwidth}
    \centering\includegraphics[width=\textwidth]{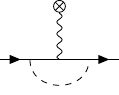}
    \caption{}
  \end{subfigure}\hfill
  \begin{subfigure}[t]{0.12\textwidth}
    \centering\includegraphics[width=\textwidth]{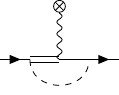}
    \caption{}
  \end{subfigure}\hfill
  \begin{subfigure}[t]{0.12\textwidth}
    \centering\includegraphics[width=\textwidth]{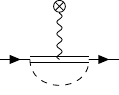}
    \caption{}
  \end{subfigure}
  \caption{Diagrams at orders $\mO(p)$, (a), and $\pthree$, (b)-(g), contributing to $F_A$. Dashed, solid single and double lines denote pions, nucleons and $\Delta$ resonances respectively; wiggly lines denote external axial fields. The open circle represents an  $\pthree$ vertex, while the rest of the vertices are $\mO(p)$. Permutations of diagrams (d) and (f) are not displayed.}
\label{fig:p3diagrams}
\end{figure}

\begin{figure}[h]
\begin{subfigure}[t]{0.12\textwidth}
\addtocounter{subfigure}{7}
    \centering\includegraphics[width=\textwidth]{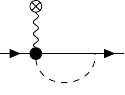}
    \caption{}
  \end{subfigure}\hfill
  \begin{subfigure}[t]{0.12\textwidth}
    \centering\includegraphics[width=\textwidth]{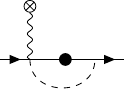}
    \caption{}
  \end{subfigure}\hfill
  \begin{subfigure}[t]{0.12\textwidth}
    \centering\includegraphics[width=\textwidth]{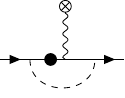}
    \caption{}
  \end{subfigure}\hfill 
  \begin{subfigure}[t]{0.12\textwidth} 
    \centering\includegraphics[width=\textwidth]{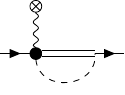}
    \caption{}
  \end{subfigure}\hfill
  \begin{subfigure}[t]{0.12\textwidth}
    \centering\includegraphics[width=\textwidth]{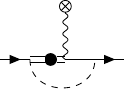}
    \caption{}
  \end{subfigure}\hfill
  \begin{subfigure}[t]{0.12\textwidth}
    \centering\includegraphics[width=\textwidth]{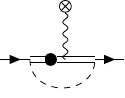}
    \caption{}
  \end{subfigure}\hfill
  \begin{subfigure}[t]{0.12\textwidth}
    \centering\includegraphics[width=\textwidth]{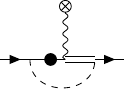}
    \caption{}
  \end{subfigure}\hfill
  \begin{subfigure}[t]{0.12\textwidth} 
    \centering\includegraphics[width=\textwidth]{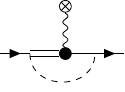}
    \caption{}
 \end{subfigure}
  \caption{$\pfour$ diagrams contributing to $F_A$. Line styles have the same meaning as in Fig.~\ref{fig:p3diagrams}. Filled circles denote  $\mO(p^2)$ vertices. Permutations of all these diagrams are properly taken into account but not explicitly represented. 
}
\label{fig:p4diagrams}
\end{figure}

The set of Feynman diagrams that contribute to the axial form factor is shown in Figs.~\ref{fig:p3diagrams} and \ref{fig:p4diagrams}. The LECs associated with each of these diagrams are listed in Table \ref{tab:diagLECs}.
At $\mO(p^4)$, there are contributions from $\mO(p^2)$ vertices but also baryon ($N, \Delta$) mass insertions. These mass insertions are treated perturbatively, i.e. we evaluate directly diagrams (i), (j), (l)-(n) of Fig.~\ref{fig:p4diagrams} without Dyson re-summations to all orders in the propagators, as is done in Ref.~\cite{Ando:2006xy}. An alternative choice has been considered in Ref.~\cite{Lutz:2020dfi}.

\begin{table}[h!]
\centering
  \begin{tabular}{|c|c|c|c|}
  \hline
      Diagrams & $\mO(p)$ & $\mO(p^2)$ & $\pthree$ \\
    \hline
    (a), (c), (d) & $\g$ & - & - \\
    (b) & - & - & $d_{16,22}$ \\
    (e) & $\g$ & - & - \\
    (f) & $\g$, $h_A$ & - & - \\
    (g) & $g_1$, $h_A$ & - & - \\
    (h) & $\g$ & $c_{2-4}$ & - \\
    (i) & $\g$ & $c_{1}$ & - \\
    (j) & $\g$ & $c_{1}$ & - \\
    (k) & $h_A$ & $b_{1,4,5}$ & - \\
    (l) & $\g$, $h_A$ & $a_1$ & - \\
    (m) & $g_1$, $h_A$ & $a_1$ & - \\
    (n) & $\g$, $h_A$ & $c_1$ & - \\
    (o) & $\g$, $h_A$ & $b_{2}$ & - \\
    wfr & $\g$, $h_A$  & $c_1$, $a_1$ & - \\
    \hline
  \end{tabular}
\caption{LECs introduced by Feynman diagrams in  Figs. \ref{fig:p3diagrams},  \ref{fig:p4diagrams} and by wave-function renormalization (wfr).}
\label{tab:diagLECs}
\end{table}
Although not diagrammatically represented, nucleon wave-function renormalization is taken into account in the standard way, as explained, for instance, before Eq.~(8) of Ref.~\cite{Alvarado:2021ibw}.
\par 
Due to the presence of baryons in the loops, the power counting of such diagrams is broken. To restore it, we subtract the power counting breaking terms (PCB) by redefining the LECs. This finite renormalization, developed in Ref.~\cite{Fuchs:2003qc}, is called extended on mass shell (EOMS). The lengthy LEC's shifts are  provided by a Mathematica notebook as supplementary material.  Notice that, as in previous works~\cite{Yao:2016vbz,Yao:2017fym,Alvarado:2021ibw}, PCB terms are identified and subtracted in an expansion in powers of $\mpi$ and $t$ but not in $\delta$.
\par 
We therefore obtain that the axial form factor within the EOMS renormalization scheme up to $\pfour$ with explicit $\Delta$ has the following structure, with superindices indicating the chiral order\footnote{In Eq.~\eqref{eq:FAexp} one might notice the following nomenclature issue: the calculation of an $\mO(p)$ diagram such as (a) in Fig.~\ref{fig:p4diagrams} yields an $\mO(p^0)$ contribution to the form factor, $\g$ in this case. This is due to the fact that the external field $a_\mu$ enters in $\mL_{\pi N}^{(1)}$ together with the partial derivative in the chiral vielbein, and as a consequence the form factor has a chiral order $p$ factorized out. When discussing the order of the calculation, we always refer to 
the power counting of the diagram not to the power of the contribution to $F_A$.}:
\bea \label{eq:FAexp}
    F_A &=& 
    \g+4 d_{16}\mpi^2+d_{22}t+F_{A\rm{(loop)}}^{(3)\slashed{\Delta}}(\g;\mpi,t)+F_{A\rm{(loop)}}^{(3)\Delta}(\g,h_A,g_1;\mpi,t)\no\\
    && +F_{A\rm{(loop)}}^{(4)\slashed{\Delta}}(\g,c_1,c_2,c_3,c_4;\mpi,t)+F_{A\rm{(loop)}}^{(4)\Delta}(\g,h_A,g_1,c_1,a_1,b_1,b_2,b_4,b_5;\mpi,t)+\mO(p^5)\,.
\eea
It is apparent that $\g$ corresponds to $F_A(0)$ in the chiral limit, while $d_{16}$ and $d_{22}$ drive the pion mass and $t$ dependencies, respectively. 
Higher order contributions are maintained to preserve the analytic structure of the loops. The $\pthree$ part of $F_A$ is given in  Eqs.~(A4-5) of Ref.~\cite{Yao:2017fym}. The $\pfour$ contribution from wave-function renormalization is given in Eq. (B1) of Ref.~\cite{Alvarado:2021ibw}. The long expression of the rest of the $\pfour$ contribution, which depends on several LECs, is also provided in a Mathematica notebook as supplementary material. As explained in Ref. \cite{Alvarado:2021ibw}, $\pfour$ $c_3$ and $c_4$ LECs enter in the combination $\widetilde{c}_4=c_4-c_3/2$, while $c_2$ enters only at $\mO(p^5)$.  Furthermore, after expanding in $\mpi$, the combination of LECs $b_4$ and $b_5$, which actually enters at $\pfour$, is $\bfour=b_4+ (12/13)\, b_5$.
  At $\mO(p^3)$ we reproduce the results of Eq.~(A4-5) of Ref.~\cite{Yao:2017fym}, except for a global factor of 36 in $\rA$\footnote{While the analytic expression in the Appendix of Ref.~\cite{Yao:2017fym} omits this factor of 36, the numerical values and plots correctly incorporate it.}, as pointed out in Ref.~\cite{Lutz:2020dfi}. However, the sign of the pion tadpoles in~\cite{Yao:2017fym} is correct, contrary to what is stated in Ref.~\cite{Lutz:2020dfi}. The reason for the confusion is probably the different sign in the definitions of the loop integral.
  
\section{\label{sec:FAfit} Analysis of LQCD results and LEC determination}
\subsection{\label{subsec:FAdata}Data set and fit strategy}
The axial form factor has long been recognized as a challenging quantity to compute in lattice QCD. However, substantial progress has been made in recent years, particularly in the treatment of excited-state contamination.
For our analysis, we adopt a combined set of lattice data for $F_A$, which includes results that are particularly careful in determining such a contamination. We take the data from the following works\footnote{Evidence of the suitability of SU(2) ChPT to study LQCD results with more than two dynamical fermions is presented in Ref.~\cite{Ren:2016aeo}.}: RQCD \cite{RQCD:2019jai}~\footnote{From Ref.~\cite{RQCD:2019jai}, only data with $m_s\sim m_{s\rm{(phys)}}$ are considered, as these are the only data suitable for an SU(2) ChPT analysis.}, NME \cite{Park:2021ypf}~\footnote{From Ref.~\cite{Park:2021ypf}, only the results of the fit strategy labeled as $\{4^{N \pi}, 3^*\}$, used to control excited-state contamination, are considered, averaging over the two renormalization methods $Z_1$ and $Z_2$.}, Mainz \cite{Capitani:2017qpc}~\footnote{From Ref.~\cite{Capitani:2017qpc} we take only data obtained with their preferred two-state-fit method.}, PNDME \cite{Jang:2023zts},  PACS \cite{Tsuji:2023llh} and ETMC \cite{Alexandrou:2020okk}  (PACS and ETMC provide results only at the physical point). This collection of data points is plotted in Fig.~\ref{fig:FADdataandDlessp3fit}(a) as a function of $Q^2$. The choice of range in $Q^2$ and $\mpi$ is addressed below.  
\par 
The data we fit have not been extrapolated in $Q^2$, $\mpi$ or to the continuum. Furthermore, we restrict ourselves to large ensembles, with $\mpi L\geq 3.5$ and neglect finite volume effects\footnote{Possible non-negligible finite-volume effects on the axial form factor have been investigated in \cite{Hermsen:2025vds} and deserve future consideration.}.
Whenever possible,  we correct for the lattice-spacing ($a$) effects by incorporating the following dependence on $a$:
\begin{equation}\label{eq:acorr}
    F_A(t_i,\mpi^i,a_i)=F_A(t_i,\mpi^i)+ (x_j+t_i y_j) a_i^{n_j} \ ,
\end{equation}
where $i$ labels single LQCD data points; $x$ and $y$ are free parameters; $j$ is the label of the lattice action and $n$ is the exponent of the corresponding discretization correction. In ChPT, the structure of linear terms in $a$ has been disclosed in Ref. \cite{Beane:2003xv} for $g_A$. In a phenomenological way, we 
further add higher order $t a$ monomials, as well as $a^2$ and $t a^2$ ones for actions with leading $a^2$ dependence. Specifically, $j=\{~\text{Mainz}$, RQCD, NME, PNDME$\}$, with $n=1$ for NME and PNDME, while $n=2$ for Mainz and RQCD. A correction for lattice-spacing dependence is not possible for PACS and ETMC, as the corresponding simulations were performed at only one lattice spacing. According to the respective publications~\cite{Tsuji:2023llh,Alexandrou:2020okk} (and our analysis of lattice-spacing effects of other data) the systematic uncertainty from this dependence should not be significant for our fit. As found in our analysis of $g_A(\mpi)$~\cite{Alvarado:2021ibw}, we anticipate that these corrections do not substantially affect the extracted physical quantities and LECs, but only  reduce the $\chi^2/{\rm dof}$.

Different low energy constants appear in the form factor, some of which are well known and can be reliably fixed, while others are left as free fitting parameters. In order to improve the fit and reduce correlations~\cite{Wesolowski:2015fqa}, we introduce naturalness Gaussian priors, $\chi^2_{\rm prior}$, for all free LECs (but $\g$), as explained in Ref.~\cite{Alvarado:2021ibw}, Eq.~(13). The breakdown scale is set to $\Lambda = 1$~GeV$\sim \, 4 \pi F_0$~\cite{Manohar:1983md,Scherer:2012xha}. As in Ref~\cite{Alvarado:2021ibw}, we foresee that a prior on $\g$ is superfluous, since its value is inescapably driven to a natural one by low $\mpi$ LQCD results for $g_A$.
 
Given that we are working in perturbation theory, the calculation has an uncertainty associated with the truncation of the chiral series. We estimate it by the difference of the orders of the calculation as proposed in Ref.~\cite{Epelbaum:2014efa}. 
 Let $X$ be an observable with a chiral expansion
\be\label{eq:X}
X = X^{(0)} + \sum_{m=1}^\infty \Delta X^{(m)} \ ,
\ee 
where $\Delta X^{(m)} = X^{(m)} - X^{(m-1)}$ refers to all the monomials that start at order $m$. If $X$ is computed up to order $n$, $X \approx X^{(n)}$, assuming that the truncation error is dominated by order $n+1$, its contribution $\Delta X^{(n+1)}$ can be estimated conservatively as~\cite{Epelbaum:2014efa,Siemens:2016hdi}
\begin{equation}
\label{eq:DX}
    |\Delta X^{(n+1)}| = \max\left\{\e^{n+1}|X^{(0)}|, \e^n|\Delta X^{(1)}|,..., \e |\Delta X^{(n)}| \right\} \,,
\end{equation}
where $\e$ denotes a chiral expansion parameter in general. Adapting the strategy of Ref.~\cite{Epelbaum:2014efa} to the FF, we adopt $\epsilon^2\equiv\max \{\mpi^2,-t\}/\Lambda^2$ as the expansion parameter. Therefore, for the $\pfour$ calculation, the truncation error is given by:
\begin{equation}\label{eq:errp5}
\Delta F_{A \chi}^{(5)}=\max\left\{\epsilon^4 \g, \epsilon^2 |\Delta F_A^{(3)}|, \left(\frac{\mpi}{\Lambda}\right) |\Delta F_A^{(4)}| \right\}\sim  \epsilon^4=\mO(p^5) \ .
\end{equation}

We also study the agreement of the leading one loop calculation $\pthree$. The truncation error in such fits is defined without implementing any information from the $\pfour$ computation, and reads
\begin{equation}
\Delta F_{A \chi}^{(4)}=\max\left\{\left(\frac{\max\{\mpi^3, -\mpi t \}}{\Lambda^3}\right) \g, \left(\frac{\mpi}{\Lambda}\right) |\Delta F_A^{(3)}| \right\}\sim \epsilon^3=\pfour \ .
\end{equation}

Overall, our $\chi^2$ is 
\be \label{eq:chi2}
\chi^2=\sum_{i} \frac{\left( F_A (t_i,M^i_\pi,a_i)- F^{i}_A \right)^2}{(\Delta F^{i}_A)^2} + \chi^2_{\rm prior} \,,
\ee
where the truncation error is added in quadrature to the LQCD one:
\be
\label{eq:toterror}
(\Delta F^{i}_A)^2  =  (\Delta F^{i}_{A \rm{LQCD}})^2 + (\Delta F_{A\chi}(t_i,\mpi^i))^2 \ . 
\ee
This procedure assigns larger uncertainties to points at high $\e$, where the convergence of the chiral expansion is poorer, reducing their impact on the fit. We perform the fits iteratively, evaluating the $\Delta F_{A\chi}$ with the LECs of the previous iteration until convergence is reached ($\Delta F_{A\chi}=0$ in the first step). 
With this method, truncation and LQCD errors are not independent, and there is no clear way to combine them into a well defined single uncertainty. For this reason, following Ref.~\cite{Siemens:2016hdi}, we plot them as different error bands for $F_A$, $g_A$ and $\rA$. In addition, as a complementary measure of the agreement of our fit to the LQCD data, we also quote  
\begin{equation}
\label{eq:chi20}
    \chi^2_0= \sum_{i} \frac{\left( F_A (t_i,M^i_\pi,a_i)- F^{i}_A \right)^2}{(\Delta F^{i}_{A \rm{LQCD}})^2} \,,
\end{equation}
without $\Delta F_{A\chi}$ or priors.

\begin{figure}[h!]
\centering
\includegraphics[width=0.36\textwidth]{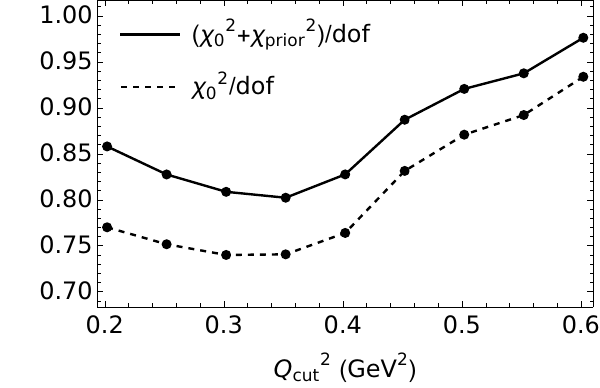}
\caption{Goodness of the fit to $F_A$ LQCD data as a function of $Q^2_{\rm cut}$: $(\chi_0^2+\chi_{\rm prior}^2)/{\rm dof}$ and  
$\chi_0^2/{\rm dof}$ are shown as solid and dashed lines respectively. See details in the text.}
\label{fig:chi2}
\end{figure}

For the purpose of establishing the range in $\mpi$ and $Q^2$ for the analysis, we search for a plateau in $\chidof$. Truncation errors imply lower weights for points at higher $\mpi$ or $Q^2$. Therefore, the question of the fit range is directly related to this uncertainty. We set $M_{\pi\rm{cut}}=402$ MeV as in our analysis of $g_A(\mpi)$~\cite{Alvarado:2021ibw} and determine $Q^2_{\rm cut}=0.36$ GeV$^2$ from the $\chidof$ plateau (Fig.~\ref{fig:chi2}), similarly to Ref.~\cite{Yao:2017fym}. These ranges are determined from the analysis of the $\pfour$ theory with explicit $\Delta$, and are retained for the simpler realizations to enable a direct comparison.

Next, the results of the ChPT fits without and with explicit $\Delta$ are presented for $\mO(p^3)$ and $\mO(p^4)$.

\subsection{\label{subsec:withoutD}$\slashed{\Delta}$ case}
\subsubsection{$\pthree$}

\begin{figure}[h!]
\centering
\begin{subfigure}[t]{0.45\textwidth}
    \includegraphics[width=\textwidth]{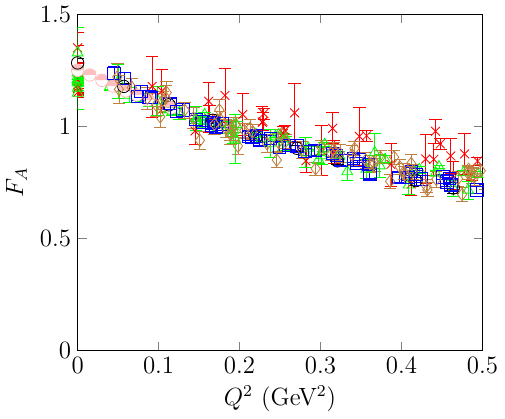}
    \caption{}
  \end{subfigure}
  \hspace{0.7cm}
  \begin{subfigure}[t]{0.45\textwidth}
    \includegraphics[width=\textwidth]{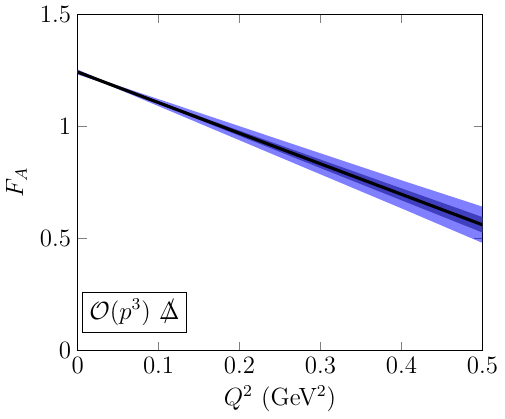}
    \caption{}
  \end{subfigure}
\caption{ {\it Left panel:} Lattice $F_A(Q^2)$ ensembles used in the fit  (without continuum extrapolation and at different $\mpi$). Red crosses, green triangles, pink half-filled circles, black circles, blue squares and brown diamonds correspond to Mainz~\cite{Capitani:2017qpc}, RQCD~\cite{RQCD:2019jai}, PACS~\cite{Tsuji:2023llh}, ETMC~\cite{Alexandrou:2020okk}, NME~\cite{Park:2021ypf} and PNDME~\cite{Jang:2023zts} respectively. {\it Right panel:} our $F_A(Q^2)$ $\mO(p^3)$ $\slashed{\Delta}$ fit at the physical point. The gray (dark) band is obtained by propagating the uncertainties in the LECs. The blue band represents truncation uncertainties $\Delta F_{A\chi}(Q^2,M_{\pi{\rm phys}})$.}
\label{fig:FADdataandDlessp3fit}
\end{figure}

At $\pthree$ without $\Delta$ there are only three free LECs: $\g$, $d_{16}$ and $d_{22}$. Their fitted values, together with those obtained for $x_j,y_j$, parametrizing the $a$ dependence,  are given in the Appendix (Table \ref{tab:FAfitLECs}). The resulting $F_A(Q^2)$, evaluated at the physical value of the pion mass, $F_A(Q^2, M_\pi = M_{\pi{\rm phys}})$, is shown in Figs.~\ref{fig:FADdataandDlessp3fit}. The lattice points at various (mostly unphysical) pion masses are shown separately. Although the plots extend to $Q^2=0.5$ GeV$^2$, we recall that the fit encompasses lattice data up to $Q^2_{\rm cut}=0.36$ GeV$^2$. For this theory and chiral order, $F_A(Q^2, M_{\pi{\rm phys}})$ exhibits a linear behavior, in agreement with the findings of Ref. \cite{Yao:2017fym}, although the lattice points exhibit some curvature. The truncation error increases with $Q^2$, as expected, and becomes larger than the one associated with the uncertainties in the LECs already at $Q^2 \gtrsim 0.1$ GeV$^2$.     

\begin{figure}[h!]
\centering
\begin{subfigure}[t]{0.45\textwidth}
    \includegraphics[width=\textwidth]{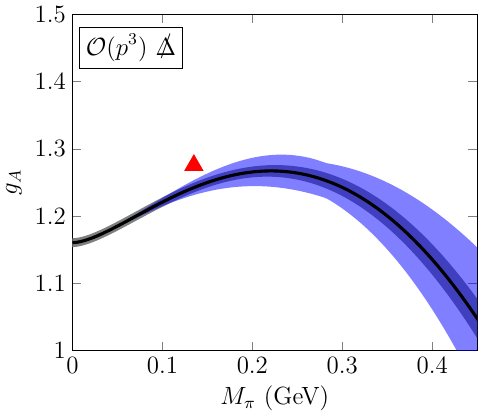}
   \caption{}
  \end{subfigure}
  \hspace{0.7cm}
  \begin{subfigure}[t]{0.47\textwidth}
    \includegraphics[width=\textwidth]{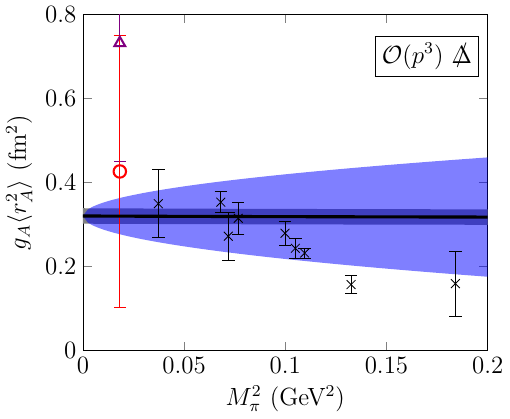}
    \caption{}
  \end{subfigure}
\caption{Pion mass dependence of $g_A$ (left) and $g_A \langle r_A^2\rangle$ (6 times the slope of $F_A$ at $Q^2=0$).  Bands have the same meaning as in Fig. \ref{fig:FADdataandDlessp3fit}. The red triangle in the left panel indicates the experimental value of $g_A$~\cite{Gorchtein:2021fce}. In the right panel, the black points are $g_A \rA$ values obtained in Ref. \cite{Capitani:2017qpc} by performing a z-expansion fit with $k_{\rm max}=1$ to their two-state-fit-method data. Notice that these points are displayed for illustration purposes only; they are not part of the analysis data-set. At the physical $\mpi$, we show the results of a recent analysis of experimental data for quasielastic neutrino scattering on hydrogen (violet triangle) and deuterium (red circle) using a z-expansion with $k_{\rm max}=6$ to parametrize $F_A(Q^2)$~\cite{MINERvA:2025ygc}.}
 \label{fig:gAandrAp3Dless}
\end{figure}

In the left panel of Figs.~\ref{fig:gAandrAp3Dless}, the pion mass dependence of $F_A(0)$ is displayed. At the physical point, the uncertainty is already dominated by the truncation error. On the right, we have chosen to show the $\mpi^2$ dependence of $g_A \langle r_A^2\rangle$, corresponding to the slope of $F_A(Q^2)$ (up to a factor 6), which can be unambiguously obtained from the matrix element of the axial current (in contrast, $\rA$ involves the ratio of the slope divided by $g_A$, also obtained perturbatively). In the present case, $g_A\rA$ is constant in $M_\pi^2$, although LQCD extractions obtain a monotonic descent. To illustrate this behavior of lattice data, the $g_A \rA$ values obtained in Ref.\cite{Capitani:2017qpc} by performing a z-expansion fit with $k_{\rm max}=1$ are displayed, although one must stress that such extrapolated points are not part of our data set.  
As can be seen in the plot, the truncation error dominates starting from right above the chiral limit. Actually, the impact of the truncation uncertainty is evident in this fit: this error assigns a lower weight to the high $\mpi$ points and, therefore, the value of $g_A \rA$ obtained with truncation error is higher than the one without it. 
 
 In any case, as we pointed out in Ref. \cite{Alvarado:2021ibw}, the main issue of the $\pthree$ description is that the $\pfour$ contribution to $g_A$ at $\mpi \gtrsim 200$ MeV turns out to be larger than the truncation uncertainty estimate of the $\pthree$ fit. This means that the truncation error and fit uncertainties are underestimated. In other words, the convergence range of this calculation is too small for the available LQCD data.

\subsubsection{$\pfour$}
At $\pfour$, $\mO(p^2)$ LECs $c_{1-4}$ enter the calculation. We fix them to the values obtained in the analysis of elastic and inelastic pion-nucleon scattering of Ref.~\cite{Siemens:2017opr} (last column of Table~3 therein). However, as explained in Ref.~\cite{Alvarado:2021ibw}, the $\pfour$ $\slashed{\Delta}$ calculation fails to describe the light-quark mass dependence of $g_A$. Therefore, we do not report a fit to the form factor at this order.

\subsection{\label{subsec:withD} $\Delta$ case}      
\subsubsection{$\pthree$}

At $\pthree$, the inclusion of the $\Delta$ introduces two additional parameters, $h_A$ and $g_1$, also at $Q^2=0$, {\it ie}, the $Q^2$ dependence does not introduce new LECs. LEC $h_A$ is fixed to its large-$N_c$ value $h_A=\frac{3 g_A}{2\sqrt{2}}\simeq1.35$ \cite{Pascalutsa:2005nd}. In the large-$N_c$ limit $\abs{g_1}=\frac{9}{5}g_A \simeq 2.29$~\cite{Hemmert:1997ye}. We adopt the large-$N_c$ value for $h_A$, whereas we do not fix $g_1$. LEC $g_1$ appears as an important coefficient for $\rA$ in the chiral limit, being strongly correlated with $d_{22}$ [Eq.~\eqref{eq:FAexp}]\footnote{LEC $g_1$ also appears in $g_A$ at $\mO(\mpi^2)$ but with a much smaller coefficient. To be precise, the impact of $g_1$ in $g_A$ is almost negligible, so it does not strongly affect the extraction of $d_{16}$ \cite{Alvarado:2021ibw}.}. 
 Actually, the main effect of leaving $g_1$ free is the large error of $d_{22}$ that results from the fit. We therefore choose to focus on the extraction of the axial radius without imposing additional assumptions on the LECs $g_1$ and $d_{22}$, 
 whose individual values are only loosely constrained, in part because of the imposed naturalness priors.
 
\begin{figure}[h!]
\centering
\begin{subfigure}[t]{0.45\textwidth}
    \includegraphics[width=\textwidth]{Fig/FAlatt26-figure0.pdf}
    \caption{}
  \end{subfigure}
  \hspace{0.7cm}
  \begin{subfigure}[t]{0.45\textwidth}
    \includegraphics[width=\textwidth]{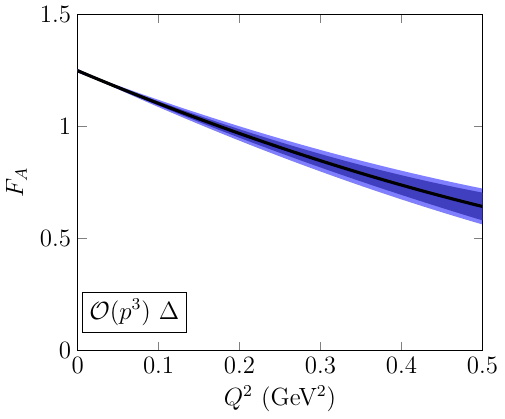}
    \caption{}
  \end{subfigure}
\caption{{\it Left panel:} Lattice $F_A(Q^2)$ ensembles used in the fit  (without continuum extrapolation and at different $\mpi$). See the caption of Fig.~\ref{fig:FADdataandDlessp3fit} for details. {\it Right panel:} $F_A(Q^2)$ at $\mO(p^3)$ with explicit $\Delta$ at the physical point. The $\chi^2$ error (gray) and the pure truncation one (blue) largely overlap.}
\label{fig:FADp3}
\end{figure}

The results of this fit are displayed in Figs.~\ref{fig:FADp3} and \ref{fig:gAandrAp3D}. The values of LECs and parameters can be found in Table \ref{tab:FAfitLECs} of the Appendix. 
As shown in Ref. \cite{Yao:2017fym}, the $\Delta$ is important to account for both the curvature of $F_A(Q^2)$\footnote{In the  $\slashed{\Delta}$ calculation of Ref. \cite{Schindler:2006it}, the curvature of $F_A(Q^2)$ originates from the explicit inclusion of the axial-vector meson $a_1(1260)$ [see Fig. (8) of that reference].} [Fig \ref{fig:FADp3} (a)] and the slope of $g_A \langle r_A^2\rangle$ as a function of $\mpi^2$ [Fig \ref{fig:gAandrAp3D} (b)].  
Unfortunately, as in the $\pthree$ $\slashed{\Delta}$ case, this description is misleadingly accurate: the $\Delta$ $\pthree$ fit underestimates the truncation error, especially in $g_A(\mpi)$. Therefore, we now focus on the $\pfour$ calculation with explicit $\Delta$, which yields more realistic results.
\begin{figure}[h!]
\centering
\begin{subfigure}[t]{0.45\textwidth}
    \includegraphics[width=\textwidth]{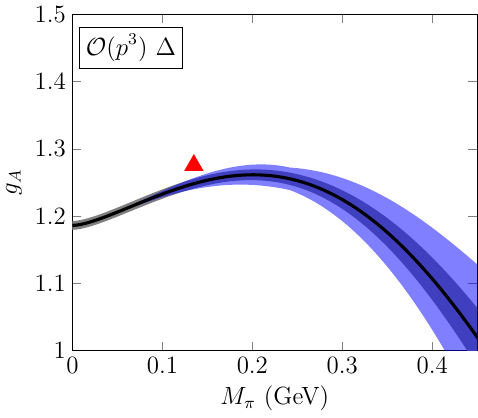}
    \caption{}
  \end{subfigure}
  \hspace{0.7cm}
  \begin{subfigure}[t]{0.47\textwidth}
    \includegraphics[width=\textwidth]{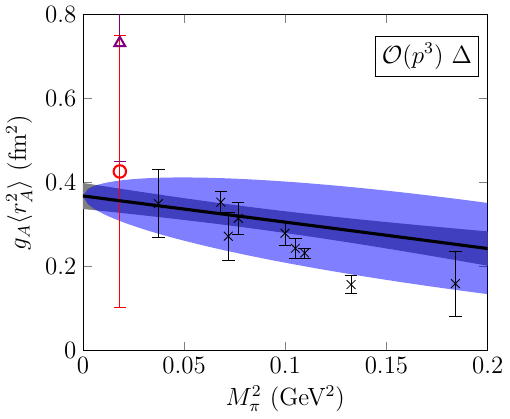}
    \caption{}
  \end{subfigure}
\caption{Results at $\mO(p^3)$ in the theory with explicit $\Delta$. The notation is the same as in Fig.~\ref{fig:gAandrAp3Dless}.}
\label{fig:gAandrAp3D}
\end{figure}

\subsubsection{\label{subsubsec:p4Dfull}$\pfour$}
\begin{figure}[h!]
\centering
\begin{subfigure}[t]{0.45\textwidth}
    \includegraphics[width=\textwidth]{Fig/FAlatt26-figure0.pdf}
    \caption{}
  \end{subfigure}
  \hspace{0.7cm}
  \begin{subfigure}[t]{0.45\textwidth}
    \includegraphics[width=\textwidth]{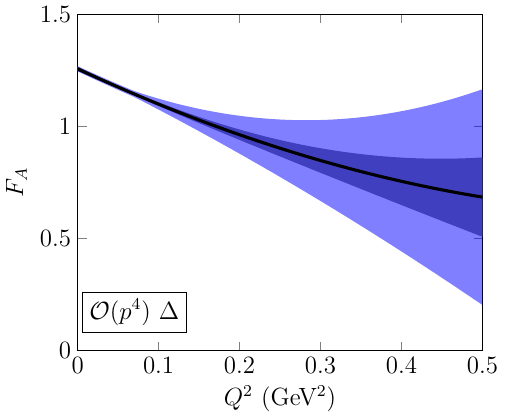}
    \caption{}
  \end{subfigure}
\caption{{\it Left panel:} Lattice $F_A(Q^2)$ ensembles used in the fit  (without continuum extrapolation and at different $\mpi$). See the caption of Fig.~\ref{fig:FADdataandDlessp3fit} for details. {\it Right panel:} $F_A(Q^2)$ at $\pfour$ with explicit $\Delta$ from the fit to the full dataset (updated). Same notation as in Fig.~\ref{fig:FADdataandDlessp3fit}(b).}
\label{fig:FADp4}
\end{figure}
\begin{figure}[h!]
\centering
\begin{subfigure}[t]{0.45\textwidth}
    \includegraphics[width=\textwidth]{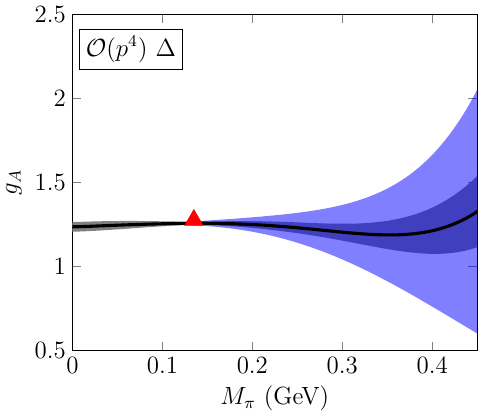}
    \caption{}
  \end{subfigure}
  \hspace{0.7cm}
  \begin{subfigure}[t]{0.47\textwidth}
    \includegraphics[width=\textwidth]{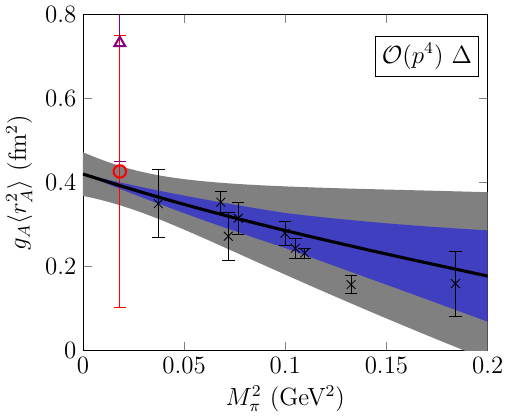}
    \caption{}
  \end{subfigure}
\caption{Results at $\mO(p^4)$ in the theory with explicit $\Delta$ from the fit to the full dataset (updated). The same notation as in Fig.~\ref{fig:gAandrAp3Dless} is used. Notice that in the radius (right panel) the error propagated from the LECs dominates over the truncation one.}
\label{fig:gAandrAp4D}
\end{figure}
At $\pfour$ with $\Delta$, we fix the $c_{1-4}$ LECs to the values extracted from $\pi N$ elastic scattering in Ref.~\cite{Siemens:2016hdi}, including the $\Delta$ resonance (all fit parameters are given in the last column of our Table~\ref{tab:FAfitLECs}). The contribution from the $\Delta$ at this order introduces the following LECs: $a_1$, $b_1$, $b_2$ and $\bfour$. We fix $a_1 = 0.90$~GeV$^{-1}$ from the LQCD $\mpi$ dependence of the $\Delta$ mass~\cite{Alvarez-Ruso:2013fza}. The combination of LECs $\bfour$ contributes to $g_A(\mpi)$ but not to the $Q^2$ dependence, playing a role in compensating the $\pfour$ $\slashed{\Delta}$ term, as discussed in~\cite{Alvarado:2021ibw}.  Like in the case of the $\pthree$ calculation, the loops with internal $\Delta$ are important to describe the $\mpi$ dependence of the axial radius. Finally, LECs $b_1$ and $b_2$ enter at $\mO(t \mpi^2)$. Their impact on the FF is limited: we only observe that the fit behaves well when these LECs take the natural values listed in Table~\ref{tab:FAfitLECs}.

The outcome of the fit is displayed in Figs.~\ref{fig:FADp4} and \ref{fig:gAandrAp4D}. 
 The low $\chi^2/{\rm dof}$ in Table~\ref{tab:FAfitLECs} is due to the relatively large truncation error at high $\mpi$, which does not strictly correspond to a $1 \sigma$ statistical deviation. In fact, by setting a lower $M_{\pi{\rm cut}}$ one obtains a higher $\chi^2/{\rm dof}$. Besides, in a naive analysis, the $\chi_0^2$ of Eq.~\eqref{eq:chi20} would have been minimized in the fit. 
 The fact that a $\chi_0^2/{\rm dof}$ close to unity is obtained, $\chi_0^2/{\rm dof}=1.31$, should be interpreted as an indication that comparing the state-of-the-art LQCD data with the $\mO(p^4)$ theory with $\Delta$ is a sensible choice in the first place. 
 Moreover, it is worth noting that the $\chi_0^2/{\rm dof}$ gets smaller as one goes from the $\mO(p^3)$ $\slashed{\Delta}$ theory to $\mO(p^3)$ with $\Delta$ and to $\mO(p^4)$ with $\Delta$ (see Tab.~\ref{tab:FAfitLECs} as well as Ref.~\cite{Yao:2017fym}). 
 
 The analysis leads to the following scenario regarding the chiral convergence:
 \begin{enumerate}
    \item As can be seen from the $M_\pi$ dependence of $g_A$ [Fig.~\ref{fig:gAandrAp4D}(a)], the $\mO(p^4)$ contribution significantly impacts the axial charge. Hence, $\mO(p^5)$, including two-loop contributions, would be required to achieve a full convergence~\cite{Bernard:2025gto}, particularly if one is interested in describing $g_A$ at relatively large $M_\pi$. This is relevant for the extraction of $d_{16}$ LEC (see the discussion below) and possibly also for the assessment of finite-volume effects. Similar conclusions were obtained in the analysis of $g_A(M_\pi)$ of Ref.~\cite{Alvarado:2021ibw}.
     \item  The plot of $F_A(Q^2,M_{\pi{\rm phys}})$ [Fig. \ref{fig:FADp4} (b)] displays an increase of the truncation error with $Q^2$, indicating that the next order can be important to describe the form factor at $\sqrt{Q^2}\gtrsim 600$ MeV. Nevertheless, the relative uncertainty introduced by the truncation is significantly smaller for the $Q^2$ dependence of the FF at $M_\pi^{\rm phys}$ compared to the $M_{\pi}$ dependence of $F_A(Q^2)$. This observation also suggests that the $\mO(p^4)$ theory with $\Delta$ is sufficiently accurate to describe the form factor at low $Q^2$.
 \end{enumerate}

 As in the $\mO(p^3)$ case, $g_A \langle r_A^2\rangle$ decreases with $M_\pi^2$ [Fig.~\ref{fig:gAandrAp4D}(b)] . Its uncertainty is larger but more realistic than the one of the $\mO(p^3)$ fit. It is dominated by the LEC uncertainty because the truncation error is relatively small. 
 
For the LEC $d_{16}$, which governs the $\mpi$ dependence of the axial charge, we obtain $d_{16}=-0.82\pm 0.72$ GeV$^{-2}$. The large uncertainty is due to the truncation error and reflects the slow convergence of $g_A(\mpi)$ in ChPT~\cite{Alvarado:2021ibw}. This determination is in agreement with the extraction from $\pi N\to \pi\pi N$ of Ref.~\cite{Siemens:2017opr}. Their value, translated to the standard EOMS scheme is $d_{16}=-1.0\pm 1.0$ GeV$^{-2}$. It is also very similar to the value obtained from our fit to $g_A$ (using a different set of LQCD ensembles) reported in Ref.~\cite{Alvarado:2021ibw}: $d_{16}=-0.88\pm 0.88$ GeV$^{-2}$. Furthermore, $\g$ and $\bfour$ are in good agreement with the values obtained in Ref ~\cite{Alvarado:2021ibw} (last column of Table 3 therein).

For $d_{22}$, which enters at tree level in the axial radius, we find $d_{22}=1.1\pm 1.4$ GeV$^{-2}$. Since this LEC is correlated with $g_1$, its uncertainty is large. The determinations of $d_{22}$ and $g_1$ in Table~\ref{tab:FAfitLECs} should be interpreted as a range of valid results; the value of one of these LECs  should not be adopted independently from the other. Nevertheless, the value $d_{22}$ is consistent with the determination from the $\pthree$ analysis of pion electroproduction of Ref.~\cite{GuerreroNavarro:2020kwb}\footnote{One has to be aware that in Ref.~\cite{GuerreroNavarro:2020kwb} the $\Delta$ is introduced only at tree level, following the so-called $\delta$-counting.}, $d_{22}=0.95\pm 0.13$ GeV$^{-2}$. The fact that we extract LECs in line with determinations from different physical processes indicates that our calculation is assessing the ChPT series in a robust way and that the convergence is good for the different observables.

Turning now to the extraction of the physical charge and radius, we first recall that the axial charge can be  determined from $\beta$ decay after radiative corrections are taken into account. For LQCD, the FLAG review reports an average of $g_A^{\text{FLAG 2+1+1}}=1.263\pm 0.010$~\cite{FlavourLatticeAveragingGroupFLAG:2024oxs}. We obtain a value of $g_A(M_{\pi\rm phys})=1.257\pm 0.011$, in good agreement with the FLAG review, and with $\sim 1.5\sigma$ tension with $g_{A} = 1.2754(13)_\mathrm{exp}(2)_\mathrm{RC}$~\cite{Gorchtein:2021fce} obtained from experiment. On the other hand, Ref. \cite{Cirigliano:2022hob} warned about unaccounted radiative corrections that might alter the comparison between isospin symmetric LQCD and experimental results. The update of Ref. \cite{Tomalak:2026wks} finds a significantly smaller value of $g_{A} = 1.240(9)$ using data-driven input. 
These values are summarized in Table~\ref{tab:gA}. Our present result is also consistent with the previous extraction we obtained using LQCD data for $g_A$ (and a different set of ensembles): $g_A(M_{\pi\rm (phys)}) = 1.260 \pm 0.012$~\cite{Alvarado:2021ibw}. Both our  determinations of $g_A$ from LQCD results have relatively small errors because the truncation uncertainty mainly affects the predictions at high $\mpi$.
\begin{table}[h!]
\centering
  \begin{tabular}{|c|c|c|c|c|}
  \hline 
       & This work & FLAG~\cite{FlavourLatticeAveragingGroupFLAG:2024oxs} & Experimental extraction~\cite{Gorchtein:2021fce} & Experimental extraction~\cite{Tomalak:2026wks}  \\
    \hline
    $g_A$ & $1.257\pm 0.011$ & $1.263\pm 0.010$ & $1.2754\pm 0.0013_{\rm exp} \pm 0.0002_{\rm RC}$ &  $1.240 \pm 0.009$ \\ 
    \hline
  \end{tabular}
  \caption{Nucleon axial charge, $g_A$, determined from the present $\mO(p^4)$ ChPT with explicit $\Delta$ meta-analysis of LQCD data, compared to the FLAG-report value and to recent extractions from experiment.}
  \label{tab:gA}
\end{table}

To extract the axial radius we take 6 times the slope of $F_A(Q^2,M_{\pi\rm phys})$ at $Q^2=0$ (the quantity in Fig. \ref{fig:gAandrAp4D} (b) at $\mpi = M_{\pi\rm phys}$) and divide it by our full result for $g_A(M_{\pi\rm phys})$. That is, we do not remove higher-order contributions in the ratio. We find $\rA(M_{\pi\rm phys})=0.312\pm 0.037$ fm$^2$ from our ChPT $\pfour$ analysis with explicit $\Delta$. The error is largely dominated by the uncertainty in the slope.   

In contrast to $g_A$, the current experimental and LQCD determinations of this quantity are less well established. In the next section, we present various available results and compare them with our own determination.

\section{\label{sec:rA} Comparison with other $\rA$ determinations}
  
\begin{figure}[h!]
\centering
\includegraphics[width=0.55\textwidth]{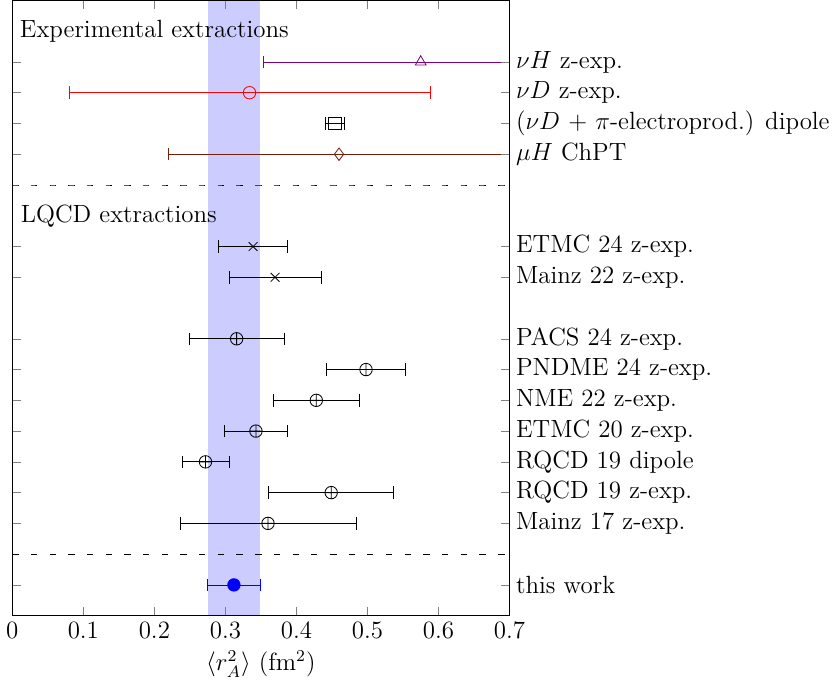}
\caption{Determinations of the physical $\rA$ from experiment and from LQCD. The experimental extractions are z-expansion fits with $k_{\rm max}=6$ to $\bar\nu H$ and $\nu D$ quasielastic scattering data \cite{MINERvA:2025ygc}, a dipole fit averaging $\nu D$ scattering and $\pi$-electroproduction~\cite{Bodek:2007ym} data, and the extraction from muonic hydrogen ($\mu H$) with ChPT input~\cite{Hill:2017wgb}. Individual LQCD values from studies  not included in our analysis are displayed as cross marks: ETMC 24~\cite{Alexandrou:2023qbg}, Mainz 22 \cite{Djukanovic:2022wru}. 
The individual extractions of the $\rA$ from LQCD works whose ensembles have been analyzed in our study, are represented by $\oplus$ symbols. These are RQCD 24~\cite{RQCD:2019jai}, NME 22~\cite{Park:2021ypf}, Mainz 17~\cite{Capitani:2017qpc}, PNDME 24~\cite{Jang:2023zts} 
PACS 24~\cite{Tsuji:2023llh} and ETMC 20~\cite{Alexandrou:2020okk}. 
Our ChPT meta-analysis of LQCD data is depicted in blue. It is worth recalling that we fit to lattice values of the form factor: the individual $\rA$ values from LQCD represented here have not been included in the fits.} 
\label{fig:rAcomparison}
\end{figure}

  The extraction of $F_A(Q^2)$ and, in particular, the radius  from experimental data is a challenging endeavor but, as outlined in Sec.~\ref{sec:FAintro} there have been recent efforts in this direction. 
  In Fig.~\ref{fig:rAcomparison} we report the values obtained in analyses of (anti)neutrino quasielastic scattering on hydrogen and deuterium, and weak muon capture in hydrogen, together with the result of our $\mO(p^4)$ ChPT with explicit $\Delta$ meta-analysis of LQCD data. In the same plot, we have listed recent LQCD determinations.\footnote{A more comprehensive account of various empirical and LQCD determinations of $\rA$ can be found in Fig. 6 of Ref. \cite{Goharipour:2025yxm}.} The plot displays the continuum and physical-point extrapolations reported by the individual LQCD studies whose data are used in our ChPT analysis. For comparison, we have also included the results of two recent LQCD determinations of $\rA$, ETMC 24~\cite{Alexandrou:2023qbg} and Mainz 22 \cite{Djukanovic:2022wru}, whose data were not available from the publications and therefore were not included in our analysis.  The figure shows that our global analysis favors lower values of $\rA$. This is also apparent from the comparison of our result with the average of recent LQCD results performed in Ref. \cite{Meyer:2026kdl}, as can be seen in Table \ref{tab:rA}, where the experimental determinations shown in Fig. \ref{fig:rAcomparison} are also quoted.  Our result is inconsistent with the higher value obtained from $\nu D$ and pion-electroproduction data using a dipole parametrization \cite{Bodek:2007ym} but is compatible with other empirical determinations that rely on the z-expansion, with much larger error bars; there is only a mild tension with the value extracted from MINERvA $\bar{\nu}H$ data in Ref. \cite{MINERvA:2025ygc}. Given the present experimental uncertainties, radiative corrections would not alter the determination of the axial radius in a statistically significant way, but they can have a noticeable impact on the form factor for $Q^2 > 0.5$ GeV$^2$ \cite{Tomalak:2026wsu}.
    
  The choice of parametrization has an impact on the extracted radius. This is clear from the difference in the predictions by the dipole parametrization in comparison to the, admittedly better motivated from the physical point of view, z-expansion,  observed in the extraction of $\rA$ from $\nu D$ data but also from the RQCD simulation. Furthermore, although the z-expansion should converge as the order of the expansion $k$ increases, some dependence on $k_{\rm max}$ is present as can be seen in the summary Table X of Ref. \cite{MINERvA:2025ygc}. Our method bears the limitation in the range of $\mpi$ and $Q^2$, inherent to the perturbative approach but, in turn, provides a model-independent parametrization, deeply rooted in QCD, and an estimate for the theoretical (truncation) uncertainty.

\begin{table}[h!]
\centering
\resizebox{\textwidth}{!}{
  \begin{tabular}{|c|c|c|c|c|c|c|}
  \hline 
       & this work & $\bar{\nu} H$ z-exp.($k_{\rm max}=6$) & $\nu D$ z-exp.($k_{\rm max}=6$) & ($\nu D$ + $\pi$-electroprod.) dip. & $\mu H$ (ChPT) & LQCD z-exp avg ($k_{\rm max}=6$) \\
    \hline
    $\rA$ (fm$^2$) & $0.312\pm 0.037$ & $0.58\pm 0.22$ & $0.33\pm 0.25$ & $0.454\pm 0.013$ & $0.46\pm 0.24$ & $0.369\pm 0.030$ \\
    \hline
  \end{tabular}
  }
  
  \caption{$\rA$ determined from our global analysis of LQCD data for the nucleon axial FF using $\mO(p^4)$ ChPT with explicit $\Delta$  compared to empirical extractions and to an average of various LQCD extrapolations of $F_A$ to the physical point~\cite{Meyer:2026kdl} in the last column\footnote{The quoted value of the LQCD average corresponds to the fit to derivatives in Ref.~\cite{Meyer:2026kdl}, which is preferred over the fit to samped form factors for statistical reasons (A. Meyer, private communication).}. The extractions from experimental data are the same as in Fig. \ref{fig:rAcomparison}.}
  \label{tab:rA}
\end{table}

\section{\label{sec:FAconcl} Conclusions and outlook}
Relativisitic chiral perturbation theory up to $\pfour$ with explicit $\Delta(1232)$ has been applied to the study of the nucleon axial form factor $F_A(Q^2)$. In this way, we derive a model-independent and systematically improvable parametrization of the pion-mass dependence of $F_A(Q^2)$ in terms of low-energy constants. Some of these LECs have been taken from previous studies of pion-nucleon elastic scattering while the rest have been fitted to the results of a set of recent LQCD simulations at various light-quark masses, accounting also for the uncertainty associated with the truncation of the perturbative expansion. 

We obtain that the $\mO(p^4)$ realization of the theory with explicit $\Delta$ provides a competitive and realistic description of the axial coupling and radius at the physical point. It also describes the main features of the LQCD data for $F_A$ at $\mpi \lesssim 400$ MeV and $\sqrt{Q^2} \lesssim 600$ MeV. To the order of our calculation, the inclusion of the $\Delta$ baryon is needed to achieve this goal. In particular, it is required to describe the $\mpi$ dependence of the lattice data and the mild curvature of $F_A(Q^2)$. Although the $\mO(p^3)$ calculation already captures these features, we find that the truncation uncertainty is underestimated at this order because the $\mO(p^4)$ contribution turns out to be sizable. 

From our fit we extract an axial charge of $g_A^{\rm phys}=1.257\pm 0.011$, in good agreement with the FLAG report~\cite{FlavourLatticeAveragingGroupFLAG:2024oxs} and in mild tension with the extraction from experiment of Ref.~\cite{Gorchtein:2021fce}, although the latter could be partially due to unaccounted radiative corrections. 
We have also investigated the axial radius, a quantity which constitutes a challenge for both experiment and LQCD theory. From our fit we obtain $\rA^{\rm{phys}}=0.312\pm 0.037$ fm$^2$. This relatively small value is consistent with most of the recent determinations performed by individual LQCD studies, and also with the result of averaging various LQCD extrapolations of $F_A(Q^2)$ to the physical point \cite{Meyer:2026kdl}. In comparison with the recent determination of $\rA$ from quasielastic neutrino scattering on hydrogen and deuterium of Ref. \cite{MINERvA:2025ygc} the result of our analysis of LQCD data agrees well with the value from $\nu D$ and is in mild tension tension with one from $\bar\nu H$. These modern results from experimental data parametrize $F_A$ with the z-expansion and have rather large uncertainties. New experiments on light targets would be useful to achieve a better understanding of the nucleon axial form factor and, in particular, the axial radius.   
Among others, we determine two important LECs in our study: $d_{16}=-0.82\pm 0.72\ {\rm GeV}^{-2}$ and $d_{22}=1.1\pm 1.4$ GeV$^{-2}$, in agreement with different phenomenological determinations. The large error in $d_{22}$ is a consequence of its correlation with $g_1$, for which we find $g_1=0.40 \pm 0.46$. In fact, these values for $d_{22}$ and $g_1$ should be used together. The LEC $d_{16}$ drives the $\mpi$ dependence of the axial charge. Its large truncation error reflects the slow convergence of ChPT for $g_A(\mpi)$; this can be addressed by extending the calculation to $\mO(p^5)$, which involves two-loop contributions. Furthermore, a global analysis of LQCD results for the axial form factor combined with pion-nucleon scattering and pion-electroproduction experimental data would likely improve the overall description, reducing, in particular, the correlations among LECs. 

\begin{acknowledgments}
The authors thank Aaron Meyer for a helpful clarification and Oleksandr Tomalak for valuable comments about radiative corrections. This research has been partially supported by CIDEGENT program with Ref. CIDEGENT/2019/015, the Spanish Ministry of Science and Innovation under Grants No. FIS2017-84038-C2-1-P, PID2020-112777GB-I00 and PID2023-147458NB-C21, funded by MICIU/AEI/10.13039/\\
501100011033, by Generalitat Valenciana grant CIPROM/2023/59 and by CEX2023-001292-S also funded by MICIU/AEI.

\end{acknowledgments}

\appendix
\section{Table of LEC values}
\begin{table}[H]
\caption{LEC values for $F_A$, both fixed and fitted to LQCD data, in the four different calculations under study. The $\chi^2/\rm{dof}$ for each fit is given at the bottom; $\chi_0^2$, defined in Eq.~\eqref{eq:chi20}, does not include theoretical errors or naturalness priors.}
\centering
  \begin{tabular}{|c|c|c|c|}
  \hline
      & $\pthree$ $\slashed{\Delta}$  & $\pthree$ $\Delta$ & $\pfour$ $\Delta$\\
    \hline
    $\g$ (free) & $1.1603\pm 0.0066$ & $1.1860\pm 0.0066$ & $1.236\pm0.031 $  \\
    $d_{16}$ (GeV$^{-2}$) (free) & $-0.959\pm 0.044$  & $1.043\pm 0.054$  & $-0.82\pm 0.72$ \\
    $d_{22}$ (GeV$^{-2}$) (free) & $1.273\pm 0.078$  & $5.2\pm 1.8$ & $1.1\pm 1.4$\ \\
    $h_A$    & - & $1.35$ & $1.35$ \\
    $g_1$ (free)    & - & $-1.19\pm 0.63$ & $0.40\pm 0.46$ \\
    $c_1$ (GeV$^{-1}$)  & - & -  &  $-1.15$ \\
    $c_2$ (GeV$^{-1}$)  & - & -   &  $1.57$ \\
    $c_3$ (GeV$^{-1}$)  & - & -  &   $-2.54$ \\
    $c_4$ (GeV$^{-1}$)  & - & - & $2.61$  \\
    $a_1$ (GeV$^{-1}$)  & -  & -  & $0.90$ \\
    $b_{1}$ (GeV$^{-2}$) (free) & -  & - & $-0.3\pm 5.0$ \\
    $b_{2}$ (GeV$^{-2}$) (free)& -  & - & $2.6\pm 2.0$ \\
    $\widetilde{b}_4$ (GeV$^{-2}$) (free)  & -  & - & $-13.04\pm 0.92$ \\
    \hline
    $x_{1}$ (fm$^{-2}$)  (free)  & $-6.1\pm 5.6$  & $-2.1\pm 5.8$  & $3\pm 14$  \\
    $x_{2}$ (fm$^{-2}$) (free)    & $-5.4\pm 2.4$   &  $-3.6\pm 2.5$ & $-2.3\pm 3.8$  \\
    $x_{3}$ (fm$^{-1}$) (free)     & $-0.06\pm 0.21$  &  $0.15\pm 0.22$ &  $0.54\pm 0.40$ \\
    $x_{4}$ (fm$^{-1}$) (free)     &  $-0.33\pm 0.28$  & $-0.10\pm 0.29$  & $0.12 \pm 0.49$  \\
    $y_{1}$ (fm$^{-2}$ GeV$^{-2}$) (free)   & $-101\pm 39$  & $-63\pm 43$  & $-47\pm 102$  \\
    $y_{2}$ (fm$^{-2}$ GeV$^{-2}$) (free)    &  $-31\pm 20$  & $-16\pm 21$  &  $-12\pm 41$ \\
    $y_{3}$ (fm$^{-1}$ GeV$^{-2}$) (free)      & $-0.6\pm 1.4$   & $0.9\pm 1.5$ &  $2.9\pm 3.5$ \\
    $y_{4}$ (fm$^{-1}$ GeV$^{-2}$) (free)     & $-2.0\pm 1.6$  & $-0.5\pm 1.7$ & $0.7\pm 3.7$  \\
    \hline
    $\m$ (GeV)    & 0.874   & 0.855  &  0.855\\
    $\md$ (GeV)    & -  & 1.166 & 1.166 \\
    \hline
    $\chi^2/\rm{dof}$    & $56.98/(158-11)=0.388$  & $47.84/(158-12)=0.328$ & $20.27/(158-15)=0.142$ \\
    $\chi_0^2/\rm{dof}$    & $767.17/(158-11)=5.22$   & $419.06/(158-12)=2.87$  & $186.63/(158-15)=1.31$  \\
    \hline
  \end{tabular}
  \label{tab:FAfitLECs}
\end{table}

\bibliography{main}

\end{document}